# Technological Leapfrogging and Manufacturing Value-added in sub-Saharan African (1990-2018)

**Ojo, Segun Michael**
Department of Economics
Redeemers University, Ede
Email: ojosegunm@yahoo.com

**Ogunleye, Edward Oladipo**
Department of Economics
Ekiti State University, Ado Ekiti
Email: edward.ogunleye@eksu.edu.ng

**Abstract**

This study examines the impact of technological leapfrogging on manufacturing value-added in SSA. The study utilizes secondary data spanning 1990 to 2018. The data is analyzed using cross-sectional autoregressive distributed lags (CS-ARDL) and cross-sectional distributed lags (CS-DL) techniques. The study found that technological leapfrogging is a positive driver of manufacturing value-added in SSA. This implies that SSA can copy the foreign technologies and adapt them for domestic uses, rather than going through the evolutionary process of the old technologies that are relatively less efficient. If the governments of SSA could reinforce their absorptive capacity and beef up productivity through proper utilization of the existing technology. The productive activities of the domestic firms will stir new innovations and discoveries that will eventually translate into indigenous technology.

**Keywords**: technological leapfrogging, manufacturing value-added, production sophistication, manufacturing export intensity, and innovation.

1. **Introduction**

The latest revolution in the global manufacturing market intensifies technological leapfrogging drive in developing countries in no small measure. Technological leapfrogging is a process whereby developing countries jump over several stages to move from low technology production to highly-modern technologies production process (Bhagavan, 2011). Studies have showed that technological leapfrogging strategy does not work in the developing countries (Wang, Wei, and Wong, 2009). Part of the questions being raised is that, can developing countries be effective at creating, owning, developing, and harnessing homegrown technology? Can developing countries adapt modern technologies into innovation cycles? (Swart, 2011). The point is that developing countries like the sub-Saharan African (SSA) may need upgrading concerning the ability to initiate home-grown technology, adapt foreign technology productively, and the welfare-generating capabilities of their markets.

Some studies observed that technological leapfrogging works well for developed countries but works poorly in developing countries. However, the result is different in the mobile phone industry. In the mobile phone industry, the developing countries are found to be ahead of the developed countries due to the fierce leapfrogging by the developing countries in the mobile phone market (James, 2013). It was argued that the economic breakthrough of the Asian miracle countries was by chance and an outcome of technology and innovation policies (TIP) other than leapfrogging strategy (Cherif and Hasanov, 2019). For instance, Fong (2009) worked on technology leapfrogging for developing countries through a literature survey across different relevant areas. The study brings up the following observations; one, technology leapfrogging is not going to be an easy venture for developing countries. Developing countries should first leapfrog in wireless technology areas because it is easier done there. Wireless technologies are relatively quicker and less costly. Two, technological leapfrogging is not a short time duration adventure. It takes a long time for leapfrogging to yield dividends.

However, the question is; how prepared are the sub-Saharan African countries for successful leapfrogging? In terms of the absorptive capacity needed to grasp foreign new technology to spur domestic industrial activities (Sauter and Watson, 2008)?

2. **Empirical Studies on Leapfrogging Strategies**

Technological leapfrogging is equally very crucial to industrialization and manufacturing development in the modern economic system. There is limited empirical literature on technological leapfrogging due to data unavailability from developing countries coupled with measuring difficulties that associate with technological leapfrogging in developing countries. However, some of the successfully studies that have been carried out on the subject matter will be reviewed in this section to uncover what has been done on the issue of technology leapfrogging in developing countries.

Wang, Wei, and Wong (2009) carried a study to investigate whether leapfrogging growth strategy raises economic growth. The study involves 49 countries comprising developed and developing countries including China, using the panel regressions technique. The result revealed no significant evidence that the strategy works reliably in those countries. Swart (2011) in a study titled; 'Africa's technology futures: three scenarios. The study investigated the present and future technological strength of Africa through two critical questions; one, can sub-Saharan Africa be effective at creating, owning, developing, and harnessing homegrown technology. Two can sub- Saharan Africa adapt non-African technologies into innovation cycles. The study utilized the descriptive statistics method and the analysis revealed that the African markets need to be strengthened in terms of the welfare-generating capabilities of the markets. African governments will need to be more effective, proactive, and responsive to stop exploitative trends that characterized the external interventions that are infiltrating the African markets.

James (2013) studied the diffusion of information technology (IT) in the historical context of innovations from developed countries. The study involved selected developed and developing countries using the S-shaped curve model of diffusion of technology developed by Rogers. The analysis revealed that the model works well for developed countries, but works poorly in developing countries. The reasons are identified to include the concentration of innovations in the rich countries and the devotion of research and development (R and D) to rich rather than poor country's problems. However, the result is different in the mobile phone industry where the developing countries are found to be ahead of the developed countries due to the fierce leapfrogging by the developing countries in the mobile phone market. Cherif and Hasanov (2019) studied the principles of industrial policy. The study used descriptive statistics and observed that the economic breakthrough of the Asian miracle countries was by chance and an outcome of technology and innovation policies (TIP). The study identified three policy thrusts that worked for the Asian tigers which are; the support of domestic producers in sophisticated industries (beyond the initial comparative advantage), export orientation, and the pursuit of fierce competition with strict accountability.

Fong (2009) worked on technology leapfrogging for developing countries through a literature survey across different relevant areas. The study brings up the following observations; technology leapfrogging is not going to be an easy venture for developing countries. Rather, developing countries should first leapfrog in wireless technology areas because it is easier there. Wireless technologies are relatively quicker and less costly. Technological leapfrogging is not a short time duration of the adventure. Rather it takes a fairly long time for leapfrogging to yield dividends. The experience of the Asian tigers was called a miracle, even though it took several decades before it materialized. According to Long (2014), the economic breakthrough of China has a historical root that is not known to both outsiders and insiders. In other words, the economic benefits of leapfrogging do not materialize quickly, because it may take considerable time before materializing; that is technological leapfrogging should be approached with long-term expectations.

Davison, Vogel, Harris, and Jones (2000) also in a literature survey studied technology leapfrogging in developing countries. The study concluded that leapfrogging in the present global context and the developmental context in developing countries cannot be easy. The study further argued that technology leapfrogging can exist in developing countries, but mere leapfrogging does not guarantee economic prosperity. Rather, economic prosperity through leapfrogging will still need the support of some other factors and policy tools that are needed to create an enabling environment for the leapfrogging process to prosper in those economies.

Korpela (2018) used descriptive statistics to analyze primary and secondary data on the study titled 'can Rwanda leapfrog to the digital economy with information and communication technology (ICT) enabled development' The study observed that it is too early to judge whether long-terms growth and development can be achieved through technological leapfrogging; as most developing countries including Rwanda have just started to diversify away from natural resources and subsistence farming economy to embrace technology-driven economy. The trajectory of development in the developed economies confirms that economic activities should be concentrated on the knowledge economy.

Similarly, Miller (2001) studied leapfrogging in the context of India's information technology industry and the internet, through an extensive survey of literature and observed that India is making headway in internet technology. This was attributed to Indian government effort geared towards encouraging the development of information technology and the internet, through various incentives and exempting the industry from inhibiting regulations and controls.

In conclusion, it can be deduced from the foregoing studies on technological leapfrogging that developing countries including the sub-Saharan African countries are making headway in ICT development. This can be attributed to the fact that ICT does not require a long time for learning and the cost of ICT infrastructure is relatively lower. But, the concern is the sustainability of economic growth and development on the platform of ICT growth. However, leapfrogging in manufacturing-driven technology in the developing countries may take a considerable period to build the absorptive capacity that will enable a successful innovative technological leapfrogging. But virtually all the studies argued that proactive policy measures can expedite the adoption of the new technologies in the developing countries, without necessarily repeating the evolutionary stages of the obsolete technologies that the developed countries went through.

3. **Theoretical issue**

Economic theorists have not developed a decisive theory for technological leapfrogging, rather studies on technological leapfrogging so far are based on empirical evidences. Therefore, this study follows in their footsteps by ascertaining the impact of technological leapfrogging on manufacturing value-added in the SSA context. The basic difference between the industrialized countries and the developing countries is the technological gap. Therefore, the tendency for leapfrogging by the latecomer countries is high, due to the market share advantage, economies of scale, and the employment opportunities that associate with introduction of new technologies. Technological change and the resultant improvement in the domestic capability to produce more outputs at lower cost are the drivers of manufacturing upgrading.

Various schools of thought have made tremendous efforts to explain the place of technological innovation in manufacturing upgrading and product diversification in the developing economies (UNIDO, 2019). Authors look forward to the government to put more effort into improving the learning process and also encourages firms to prioritize R and D by devoting huge investment for R and D projects and programs. The successful economies in the world are those that restructured their productivity from low value-added production to high value-added activities (Salazar-Xirinachs et al. 2014).

4. **Method of Estimation**

The effectiveness of leapfrogging strategy at spurring manufacturing activities in sub-Saharan African countries is the focal point of our analysis in this sub-section. This analysis is based on the work of Wang, Wei and Wong (2009) which investigated the nexus between leapfrogging strategy and economic growth in China. This study adapts the study for manufacturing sector in SSA to ascertain the efficacy of technological leapfrogging strategy for industrialization and manufacturing sector development in the African sub-region. The mathematical relationship between the dependent variable and the independent variables is given as;

$$MVA_{it} = \beta_{i0} + \beta_{i1}MVA_{it-1} + \beta_{i2}LFRG_{it} + \beta_{i5}SPH_{it} + \beta_{i6}HC_{it} + \beta_{i7}GEF_{it} + \beta_{i8}FDN_{it} + \varepsilon_{it} \qquad (1)$$

Where, $i = 1, 2, 3 \ldots n$, $t = 1, 2, 3 \ldots T$. $MVA_{i,t}$ is the dependent variable, which is manufacturing value-added, $MVA_{i,t-1}$ is one year lag of the dependent variable representing the previous year's level of manufacturing value-added. The focus variable is technological leapfrogging ($LFRG_{it}$) and it is measured as share of advanced technology products in total exports (Wang, Wei and Wong, 2009). The explanatory variables include production sophistication ($SPH_{it}$), Human capital ($HC_{it}$), government effectiveness ($GEF_{it}$), and financial deepening ($FDN_{it}$). Production sophistication is captured by production sophistication index calculated as weighted average of a country's per capita GDP where the weights are manufacturing export intensity (MEI) of individual country which is calculated as the market share in world manufacturing export. It is a country's exports of manufacturing divided by the world's total exports of manufacturing (Cherif and Hasanov, 2019; Anand, Mishra, and Spatafora, 2012). Production sophistication in this study is a proxy for innovation and technological advancement that propels manufacturing value-added growth. Human capital ($HC_{it}$), human capital is used as proxy factor endowment aspect of manufacturing in the model; and government effectiveness ($GEF_{it}$) is a proxy for the government policy concerning manufacturing and leapfrogging. $\varepsilon_i$ is the error terms. However, in Wang, Wei and Wong (2009) model the financial sector is left out. Therefore, we incorporate financial deepening ($FDN_{it}$) as a control variable representing the financial sector development in the model (financial sector development is a key facilitator of manufacturing value-added).

This study will utilize cross-sectional autoregressive distributed lags (CS-ARDL) and cross-sectional distributed lags (CS-DL) techniques. The two methods are complementary in nature. In other words, CS-ARDL is suitable for short-run relationship in a heterogeneous panel while CS-DL is appropriate for estimating average long-run coefficients. The two models allow for heterogeneity among the cross-sectional units and error cross-sectional dependence in a panel framework. The cross-sectional units in this study are heterogeneous as they operate at different levels of development, technology, political system, and macroeconomic environment. The assumption of homogeneity in the context of the cross-sectional units of this study, may leave much to be desired. The likelihood of error cross-sectional dependence due to unobserved common factors is high, because the countries are exposed to common international shocks.

The traditional ARDL model assumes homogeneity among the units (countries) and cross-sectional independence of the error terms. Whereas, error terms are likely to be cross dependence due to the presence of unobserved common factors in the models. Therefore, the traditional ARDL model is modified to take care of slope heterogeneity and error cross-sectional dependence in the model. The modified version of ARDL is known as CS-ARDL which is stated as;

$$Y_{it} = \sum_{j=1}^{r} \beta_{ij} Y_{i, t-j} + \sum_{j=0}^{p} \theta'_{ij} X_{i, t-1} + \forall_i + \mu_{it} \qquad (2)$$

$$\mu_{it} = \tau'_i f_t + \varepsilon_{it} \qquad (3)$$

Where, $f_t$ is an $m \times 1$ vector of unobserved common factors; $\tau'_i$ is the corresponding factor loading and $\varepsilon_{it}$ are the idiosyncratic errors. According to Chudik, Mohaddes, Pesaran, and Raissi (2017) and Chudik, Alexander, Pesaran and Tosetti (2011) those idiosyncratic errors could be serially or cross-correlated or uncorrelated with the common factors. Unobserved common factors are the sources of error cross-section dependence and it drive the variables differently across the countries.

$$X_{i,t-1} = \pi'_i f_t + v_{it}$$

Where, $X_{i,t-1}$, are the vector of the explanatory variables. The assumption here is that, the explanatory variables could be influenced by the unobserved common factors ($f_t$), $\pi'_i$ is the nXm matrix of factor loadings and $v_{it}$ are the idiosyncratic errors that associated with $X_{i,t-1}$ which are believed to be uncorrelated with $\varepsilon_{it}$.

$$\mu_{it} = \sum_{i=1}^{mfs} \rho^s_{it} f^s_{it} + \sum_{i=1}^{mfw} \rho^w_{it} f^w_{it} + e_{it} \qquad (4)$$

According to Chudik et al. (2011), the common factor structure can be expressed as a combination of finite number ($wfs$) of strong factors ($f^s_{it}$) that can be correlated with the regressors of the original model, and a finite number ($mfw$) of weak, semi-weak and semi-strong factors ($f^w_{it}$) which may be affecting a segment of some countries among the sample. Chudik, Mohaddes, Pesaran and Raissi (2017), gave the example of strong factors or shocks to include structural changes, instance of global financial cycle changes, interest rates, and commodity price shocks.

The weak, semi-weak and semi-strong factors can be attributed to spillovers of domestic industrial activity, domestic consumption, geographical proximity, research and development, climate factor, agricultural productivity and natural resources (Eberhardt and Vollrath, 2016; Eberhardt, Helmers and Strauss, 2013 and Holly, Pesaran and Yamagata, 2010).

Chudik, Mohadders, Pesaran and Raissi (2013) proposed another technique of estimating the long-run coefficients directly known as the 'Distributed lag' (DL) method. With this approach equation 2 can be rewritten as;

$$Y_{i,t} = \delta_i X_{i,t} + \pi'_i(L)\Delta X_{i,t} + \hat{\varepsilon}_{i,t} \tag{5}$$

Where, $\hat{\varepsilon}_{i,t} = \vartheta(L)^{-1}\varepsilon_{i,t}$; $\vartheta_i(L) = 1 - \sum_{j=1}^{p}\vartheta_{i,j}L^j$; $\delta_i = \rho_i(1); \rho_i(L) = \vartheta_i^{-1}(L)\varphi_i(L) = \sum_{j=0}^{\infty}\rho_{i,j}L^j$; $\varphi_i(L) = \sum_{j=0}^{r}\varphi_{i,j}L^j$; and $\pi_i(L) = \sum_{j=0}^{\infty}\sum_{s=j+1}^{\infty}\rho_s L^j$

Obtaining a consistent estimate of the average log-run coefficients $\delta_i$ requires the following conditions; the roots of $\vartheta_i(L)$ should lie outside the unit circle. The coefficients of $\pi_i(L)$ are assumed to be exponentially decaying. And there is no reverse causality from lagged dependent variables to independent variables. After the long-run coefficients $\hat{\delta}_i$ are estimated, the values will be used to estimate the average long-run effects by averaging $\hat{\delta}_i$ across the units (i) using a simple formula;

$$\hat{\bar{\delta}}_{MG} = N^{-1}\sum_{i}^{N}\hat{\delta}_i$$

## 5. Discussion of Result

The summary of the descriptive statistics on the model is presented in table 1. The skewness results displayed in the table indicate high skewness for total employment (TE), financial deepening (FDPN), gross domestic product (GDP), gross capital formation (GCF), production sophistication (SPH), and government effectiveness (GEFF). While manufacturing value-added (DVA), human capital (HC), and technological leapfrogging (LFGN) are moderately skewed. The results for kurtosis signify negative kurtosis for manufacturing value-added (DVA), human capital (HC), and technological leapfrogging (LFGN). Total employment (TE), financial deepening (FDPN), gross domestic product (GDP), gross capital formation (GCF), production sophistication (SPH), and government effectiveness (GEFF) have positive kurtosis.

**Table 1 Descriptive Statistics on the Technological Leapfrogging Model**

|         | DVA      | SPH       | HC       | GEFF      | LFGN      | TE        | FDPN      | GDP       | GCF      |
|---------|----------|-----------|----------|-----------|-----------|-----------|-----------|-----------|----------|
| **M.**     | 13.1773  | 3.15713   | -0.63121 | -0.42632  | 5.615926  | 14.55147  | 17.55207  | 22.58437  | 7.158243 |
| **Med.**   | 13.11433 | 2.22419   | -0.61299 | -0.229457 | -1.794016 | 15.34375  | 19.96008  | 23.03534  | 6.102801 |
| **Max.**   | 18.36305 | 21.43622  | -0.06161 | 1.01277   | 27.60508  | 18.07018  | 27.01586  | 27.0492   | 24.45297 |
| **Min.**   | 9.425452 | -3.629249 | -1.39333 | -4.669132 | -13.05149 | 3.965241  | -15.99871 | 5.263015  | 1.548354 |
| **S. Dev.**| 1.845086 | 3.886616  | 0.292816 | 0.891381  | 11.17407  | 2.563124  | 6.956584  | 3.038155  | 5.117777 |
| **Skew**   | 0.425869 | 2.761977  | -0.53274 | -1.580943 | 0.637912  | -2.564656 | -2.020568 | -3.640763 | 1.772189 |
| **Kurt.**  | 2.972199 | 9.911429  | 2.888023 | 6.15751   | 1.607985  | 10.86934  | 6.713355  | 21.11738  | 5.273898 |
| **CV**     | 0.140    | 1.231     | -0.463   | -2.09     | 1.989     | 0.176     | 0.396     | 0.1345    | 0.7149   |

**Source: author's computation**

The values of the coefficient of variation (CV) in the table show that the standard deviation is not high for the majority of the variables in the model. More specifically, manufacturing value-added (DVA), human capital (HC), total employment (TE), financial deepening (FDPN), gross domestic product (GDP), and gross capital formation (GCF) are less dispersed. Production sophistication (SPH), government effectiveness (GEFF), and technological leapfrogging (LFGN) exhibit high dispersion in the model.

However, the high dispersion observed in the data sets does not constitute threat to this analysis because cross-sectional distributed lags (CS-DL) and cross-sectional autoregressive distributed lag (CS-ARDL) models account for heteroscedasticity in the residuals through the variance of the mean group and pooled estimator. This implies that CS-DL and CS-ARDL return unbiased and consistent estimators in the face of heteroscedasticity in a model.

The result of the analysis of the relationship between manufacturing value-added and technological leapfrogging is reported in table 2. The table contains two sets of results; the CS-DL and CS-ARDL results. CS-ARDL usually reports two estimations namely, the short-run coefficients and the long-run coefficients together with the speed of adjustment that is usually referred to as cointegration in the CS-ARDL context. The long-run coefficients in the CS-ARDL framework are computed outside the model's apparatus using the estimated short-run coefficients. Therefore, the long-run coefficient estimated through CS-ARDL is not a reliable long-run estimator because the model does not return a consistent estimator if heterogeneity is present in the model due to the inclusion of the lagged dependent variable. One good thing about CS-ARDL is that it estimates the speed of adjustment which reveals whether there is cointegration in the model or not.

**Table 2 Estimation Based on CS-DL Approach**

|  | CS-DL (p=1, 2, 3) | CS-ARDL |
|---|---|---|
| **Variables** | **Coefficients** | **Coefficients** |
| **LFGN** | 0.1169003*** | 0.0510871 |
|  | (0.0296537) | (0.0552834) |
| **FDPN** | 0.0222863** | 0.218844 |
|  | (0.0100793) | (0.017128) |
| **TE** | -0.6722463 | -0.394687 |
|  | (1.019763) | (2.584075) |
| **GDP** | 0.6471854*** | 0.3631497 |
|  | (0.2366478) | (0.1145543) |
| **GCF** | 0.025226** | 0.0180857 |
|  | (0.0136972) | (0.221671) |
| **SPH** | -0.0872533 | -0.0324361 |
|  | (0.0316972) | (0.1731145) |
| **HC** | 0.7174998*** | -0.5551111 |
|  | (0.2713632) | (0.5173925) |
| **GEFF** | -0.0027168 | -0.2138504 |
|  | (0.0864205) | (0.1145543) |
| **Cointegration** |  | -1.102649 |
| **CD Statistics** | 37.29 | 16.02 |
| **CD p-value** | 0.0000 | 0.0000 |
| **Obs** | 728 | 728 |

**Source: author's computation**
Note: CS-DL and CS-ARDL estimates. Standard error in parentheses, variables are measured in logs. Control for cross-sectional dependence (CSD): CCE Common Correlated Effects (for one 1 lag). ***, **, * denote significant at 1, 5 and 10% respectively.

The focus variable in this analysis is technological leapfrogging because the study seeks to ascertain the impact of technological leapfrogging on manufacturing value-added in SSA. In the CS-DL result, the coefficient of technological leapfrogging is positive (0.1169003) and significant at a 1% level of significance. This implies that technological leapfrogging spurs manufacturing productivity in SSA. More specifically, it implies that a one percent increase in technological leapfrogging increases manufacturing value-added by 0.12% in SSA. This result supports the observation of Cherif and Hasanov (2019) about the Asian Miracles who they claimed got their miracles through the support of domestic producers in sophisticated industries. This implies that the SSA does not need to repeat the process of dirty and wasteful old technologies in their bid to spur industrialization and manufacturing upgrading. This study reveals that technological leapfrogging does not hurt manufacturing productivity in SSA. Rather, the sub-Saharan African countries can plan out a proactive technological leapfrogging agenda to spur industrialization and manufacturing development in the manner of the Asian Tigger's.

Financial deepening is positive and significant in this analysis. The average long-run coefficient of the relationship between manufacturing value-added and financial deepening is estimated at '0.0222863', and it is significant at the 5% level of significance. Sines (1979) observed a similar relationship between financial deepening and manufacturing processing for Venezuela. Akinmulegun and Akinde (2019) also recorded a similar result for Nigeria. The result indicates that financial sector development and diversification is required for SSA industrial development. Gross domestic product (GDP), gross capital formation, and human capital are positively signed and statistically significant in this analysis. While total employment rate, production sophistication, and government effectiveness are negative and not significant.

In the CS-ARDL analysis, none of the variables is statistically significant. Although some are rightly signed, but none is significant perhaps due to the heterogeneity problem and endogeneity issue accentuated by the inclusion of lagged dependent variables in the model. All the CS-ARDL standard errors for the explanatory variables are higher than the corresponding standard errors in the CS-DL model. In other words, CS-ARDL estimators exhibited higher variance than CS-DL estimators which implies that CS-DL is more efficient at estimating long-run coefficients than CS-ARDL.

**Robustness Analysis**

The robustness analysis of this model is done by the addition and subtraction of variables from the original model and observe the resultant changes on the signs and magnitudes of the estimated coefficients in the model.

**Table 3 Estimates Based on CS-DL Approach**

| | CS-DL (p=1, 2, 3) | | |
|---|---|---|---|
| | Model (A) | Model (B) | Model (C) |
| **Variables** | **Coefficients** | **Coefficients** | **Coefficients** |
| **LFGN** | 0.0101346** | 0.0793217*** | 0.1169003*** |
| | (0.0054554) | (0.0328842) | (0.0296537) |
| **FDPN** | - | 0.0304547** | 0.0222863** |
| | | (.0157504) | (0.0100793) |
| **GDP** | - | 0.5159022*** | 0.6471854*** |
| | | (0.1897809) | (0.2366478) |
| **GCF** | - | 0.0225565* | 0.025226** |
| | | (0.0139247) | (0.0136972) |
| **SPH** | - | -0.0720157 | -0.0872533 |
| | | (.0354235) | (0.0316972) |
| **HC** | - | -0.3526655 | 0.7174998*** |
| | | (0.3743572) | (0.2713632) |
| **GEFF** | - | - | -0.0027168 |
| | | | (0.0864205) |
| **TE** | - | - | -0.6722463 |
| | | | (1.019763) |
| **CD Statistics** | 49.73 | 28.14 | 37.29 |
| **CD p-value** | 0.0000 | 0.0000 | 0.0000 |
| **Obs** | 702 | 728 | 728 |

**Source: author's computation**
Note: cross-Sectional Distributed Lags (CS-DL) estimates. Standard error in parentheses, variables are measured in logs. The reported coefficients are long-run parameters. Control for cross-sectional dependence (CSD): CCE Common Correlated Effects (for one 1 lag). ***, **, * denote significant at 1, 5 and 10% respectively.

Table 3 contains the results of three equations of CS-DL models with a different combination of explanatory variables. Model (A) regresses only technological leapfrogging as an independent variable on manufacturing value-added, model (B) increased the number of the explanatory variables to six including the focus variable (technological leapfrogging) and model (C) is the original model previously analyzed in table 2. Technological leapfrogging is positive and statistically significant in the three models. In other words, technological leapfrogging is positive and significant when it was regressed on MVA as a lone independent variable and it remains significant and positive when five more explanatory variables were added in the model (B). These two analyses bear out the result of the initial model where technological leapfrogging was regressed on manufacturing value-added amidst seven other explanatory variables, which is label as the model (C) in table 3. The magnitude of the estimated coefficients for the technological leapfrogging parameters ranges from 0.1169003 to 0.0101346 across the three analyses. The marginal differences in the coefficients of technological leapfrogging across the three analyses signifies the robustness of the parameters of the original model and the consistency of CS-DL estimators. Therefore, it can be concluded that technological leapfrogging promotes industrialization and manufacturing development in sub-Saharan Africa.

The robustness of this analysis is further evident by the signs and magnitudes of other explanatory variables like financial deepening, gross domestic product, and gross capital formation. Financial deepening, gross domestic product, and gross capital formation are positively signed and statistically significant in model (B) and (C). The marginal differences in the values of their coefficients also confirmed the robustness of the estimates.

## 6. Conclusion

The high wave of technological advancement that started from the mid-twentieth century till the late twenty-century tremendously transformed the global economic landscape, the domestic macroeconomic environment, and the process of carrying out economic transactions across nations. Therefore, the resultant efficiency and ease of doing business put pressure on all organizations and countries to imbibe the new initiatives. The greatest of the pressure is on the developing countries including the SSA who are behind the industrialized nations in terms of technological know-how. So, the developing countries can only avoid being left out or marginalized particularly in the international markets; if they join up with the advanced countries in the new stream of technological drive (technological leapfrogging) or start the development of their technological apparatus from scratch.

However, the argument about whether the developing countries should leapfrog or start their technological formation from scratch is still opened to debate. But this study found that technological leapfrogging is a positive driver of manufacturing value-added in SSA. This implies that SSA can copy the foreign existing technologies and adapt them for domestic uses, rather than going through the evolutionary process of the old technologies that are relatively less efficient. If the governments of SSA could reinforce their absorptive capacity and beef up productivity through proper utilization of the existing technology. The productive activities of the domestic firms will stir new innovations and discoveries that will eventually translate into indigenous technology.

**Refrences**

Akinmulegun, S, O. and Akinde, J A. (2019) "Financial deepening and manufacturing sector performance in nigeria (1981-2017). *IOSR Journal of Economics and Finance (IOSR-JEF), vol. 10, no. 4, 2019, pp. 18-27*

Anand, R., Mishra, S. and Spatafora, N. (2012) Structural transformation and the sophistication of production. *IMF Working Paper. WP/12/59*

Bhagavan, M. R. (2011) Globalization of technology – technological leapfrogging by developing countries. *Stockholm Environment Institute, Stockholm, Sweden.©encyclopedia of life support systems (eolss)*

Cherif, R. and Hasanov F. (2019) The return of the policy that shall not be named: principles of industrial policy. *IMF Working Paper*

Chudik, A., Mohaddes, K., Pesaran, M. H., and Raissi, M. (2017). "Is there a debt-threshold effect on output growth?" *Review of Economics and Statistics, 99(1):135–150.*

Chudik, Alexander, M. Hashem Pesaran and Elisa Tosetti. (2011). Weak and strong cross-section dependence and estimation of large panels. *The Econometrics Journal 14(1):C45–C90.*

Chudik, Alexander, Kamiar Mohaddes, M. Hashem Pesaran and Mehdi Raissi. 2013. "Debt, inflation and growth: robust estimation of long-run effects in dynamic panel data models." *Cambridge Working Papers in Economics Debt.*

Davison, R., Vogel D., Harris R. and Jones N.(2000) Technology leapfrogging in developing countries - an inevitable luxury? *Electronic Journal of Information Systems in Developing Countries ·*

Eberhardt, M. And Vollrath, D. (2016). 'The effect of agricultural technology on the speed of development', forthcoming in world development

Eberhardt, M., Helmers, C., And Strauss, H. (2013). 'Do spillovers matter when estimating private returns to RandD?', *The Review Of Economics And Statistics, Vol. 95, Pp. 436-448.*

Fong, M. W. L. (2009) Technology leapfrogging for developing countries. *Global Information Technology Encyclopedia of Information Science and Technology, Second Edition*

Holly, S., Pesaran, M. H., And Yamagata, T. (2010). 'A spatial-temporal model of house prices in the USA', *Journal Of Econometrics, Vol. 158, Pp. 160-173.*

James, J. (2013) The diffusion of it in the historical context of innovations from developed countries. *Springerlink. Soc Indic Res (2013) 111:175–184*

Korpela K. (2018) Can Rwanda leapfrog to the digital economy with ICT enabled development? A case study of a developing country determined to modernize without the traditional model. *Lund University School of Economic and Management*

Long V. (2014) A technological capabilities perspective on catching up: the case of the Chinese information and communications technology industry. *Industrial Economics and Management Royal Institute of Technology, KTH S- 100 44 Stockholm, Sweden*

Miller, Robert R.(2001). Leapfrogging India's information technology industry and the Internet (English). *International Finance Corporation discussion paper; no. IFD 42*IFC working paper series. Washington, D.C: The World Bank.*

Salazar-Xirinachs, Nübler and Kozul-Wright (2014). Industrial policy, productive transformation and jobs: theory, history and practice. *Text book publication*

Sauter, R. and Watson, J. (2008) Technology leapfrogging: a review of the evidence a report for DFID. *Sussex Energy Group SPRU (Science and Technology Policy Research), University of Sussex 3rd October 2008*

Sines, R. (1979). Financial deepening" and industrial production: a microeconomic analysis of the Venezuelan food processing sector. *Social and Economic Studies, 28*(2), 450-474. Retrieved September 27, 2020, from http://www.jstor.org/stable/27861758

Swart, D. (2011) Africa's technology futures: three scenarios. *The Frederick S. Pardee Center for the Study of the Longer-Range Future Boston University. The PARDee PAPeRS / No. 14 / July 2011*

United Nations Industrial Development Organization, (2019). Industrial development Report 2019. Demand for manufacturing: driving inclusive and sustainable industrial development. Vienna.

Wang Z, Wei S and Wong (2009) Does a leapfrogging growth strategy raise growth rate? Evidence across countries and within China. *Joint Symposium of U.S-China Advanced Technology Trade and Industrial Development. October 23-24, 2009*